\documentclass[11pt]{article}
\usepackage{moriond,graphicx}

\makeatletter
\newcommand{\figcaption}[1]{\def\@captype{figure}\caption{#1}}
\newcommand{\tblcaption}[1]{\def\@captype{table}\caption{#1}}

\makeatother

\def\Journal#1#2#3#4{{#1} {\bf #2}, #3 (#4)}

\def\NIMA{{\em Nucl. Instrum. Methods} A}

\def\PLB{{\em Phys. Lett.}  B}
\def\PRL{\em Phys. Rev. Lett.}
\def\PRD{{\em Phys. Rev.} D}

\def\be{\begin{equation}}
\def\ee{\end{equation}}
\def\bea{\begin{eqnarray}}
\def\eea{\end{eqnarray}}

\begin{document}
\vspace*{4cm}
\title{THE RECENT RESULTS OF THE SOLAR NEUTRINO MEASUREMENT IN BOREXINO}

\author{ Y.KOSHIO on behalf of Borexino Collaboration : \\
G. Bellini, J. Benziger, D. Bick, S. Bonetti, M. Buizza Avanzini, \\
B. Caccianiga, L. Cadonati, F. Calaprice, C. Carraro, P. Cavalcante, \\
A. Chavarria, D. DâAngelo, S. Davini, A. Derbin, A.Etenko, \\
F. von Feilitzsch, K. Fomenko, D. Franco, C. Galbiati, C. Ghiano, \\
M. Giammarchi, M. GÂ¨oger-Neff, A. Goretti, L. Grandi, E. Guardincerri, \\
S. Hardy, Aldo Ianni, Andrea Ianni, V. Kobychev, D. Korablev, G. Korga, \\
Y. Koshio, D. Kryn, T. Lewke, B. Loer, F. Lombardi, L. Ludhova, I. Machulin, \\
W. Maneschg, G. Manuzio, Q. Meindl, E. Meroni, L. Miramonti, M. Misiaszek, \\
D. Montanari, P. Mosteiro, V. Muratova, L. Oberauer, M. Obolensky, \\
M. Pallavicini, L. Papp, L. Perasso, S. Perasso, A. Pocar, R.S. Raghavan, \\
G. Ranucci, A.Razeto, A. Re, A. Romani, A. Sabelnikov, R. Saldanha, \\
C. Salvo, S. SchÂ¨onert, H. Simgen, M. Skorokhvatov, O. Smirnov, \\
A. Sotnikov, S. Sukhotin, Y. Suvorov, R. Tartaglia, G.Testera, \\
D. Vignaud, R.B. Vogelaar, J. Winter, M. Wojcik, A. Wright, M. Wurm, J. Xu, \\
O.Zaimidoroga, S. Zavatarelli, and G. Zuzel }

\address{Laboratori Nazionali del Gran Sasso, INFN, \\
S.S. 17 BIS km. 18.910, 67010 Assergi L'Aquila, Italy \\
koshio@lngs.infn.it}

\maketitle\abstracts{
The recent released results of 153.62 ton$\cdot$year exposure
of solar neutrino data in Borexino are here discussed.
Borexino is a multi-purpose detector with large volume liquid scintillator,
located in the underground halls of the Laboratori Nazionali del Gran Sasso
in Italy. The experiment is running since 2007. The first realtime
$^7$Be solar neutrino measurement has been released in 2008.
Thanks to the precise detector calibration in 2009, the $^7$Be flux measurement
has been reached with an accuracy better than 5\%. The result related to the
day/night effect in the $^7$Be energy region is also discussed.
These results validate the MSW-LMA model for solar neutrino oscillation.}

\section{Introduction}
\label{sec1}
\begin{figure}[h]
 \begin{minipage}[c]{0.48\textwidth}
 \begin{center}
  \resizebox{\textwidth}{!}{\includegraphics{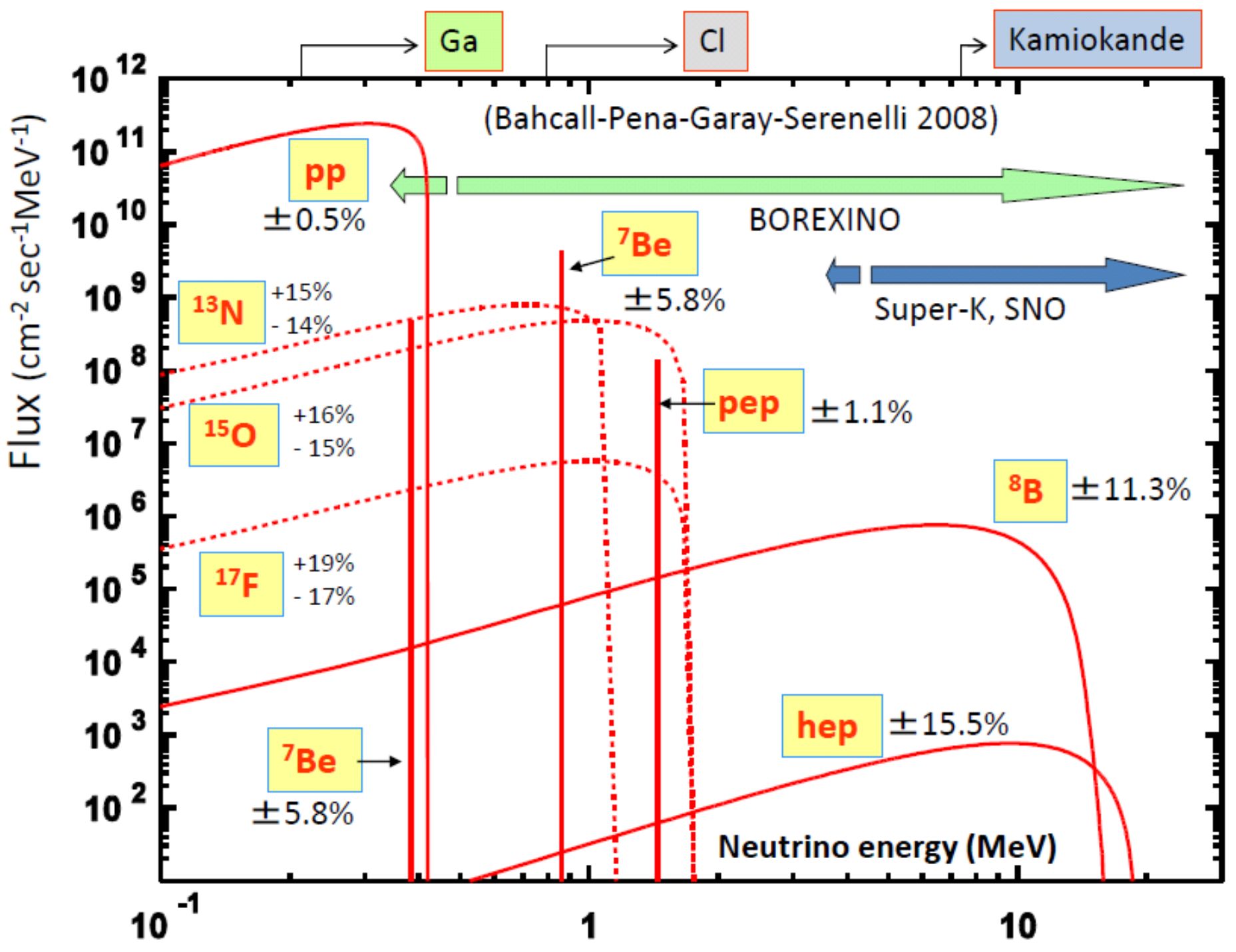}}
 \end{center}
 \caption{Solar neutrino spectra with sensitive energy region in each experiment.}
 \label{fig:nuspc}
 \end{minipage}
 \hfill
 \begin{minipage}[c]{0.48\textwidth}
  \begin{center}
  \resizebox{0.85\textwidth}{!}{\includegraphics{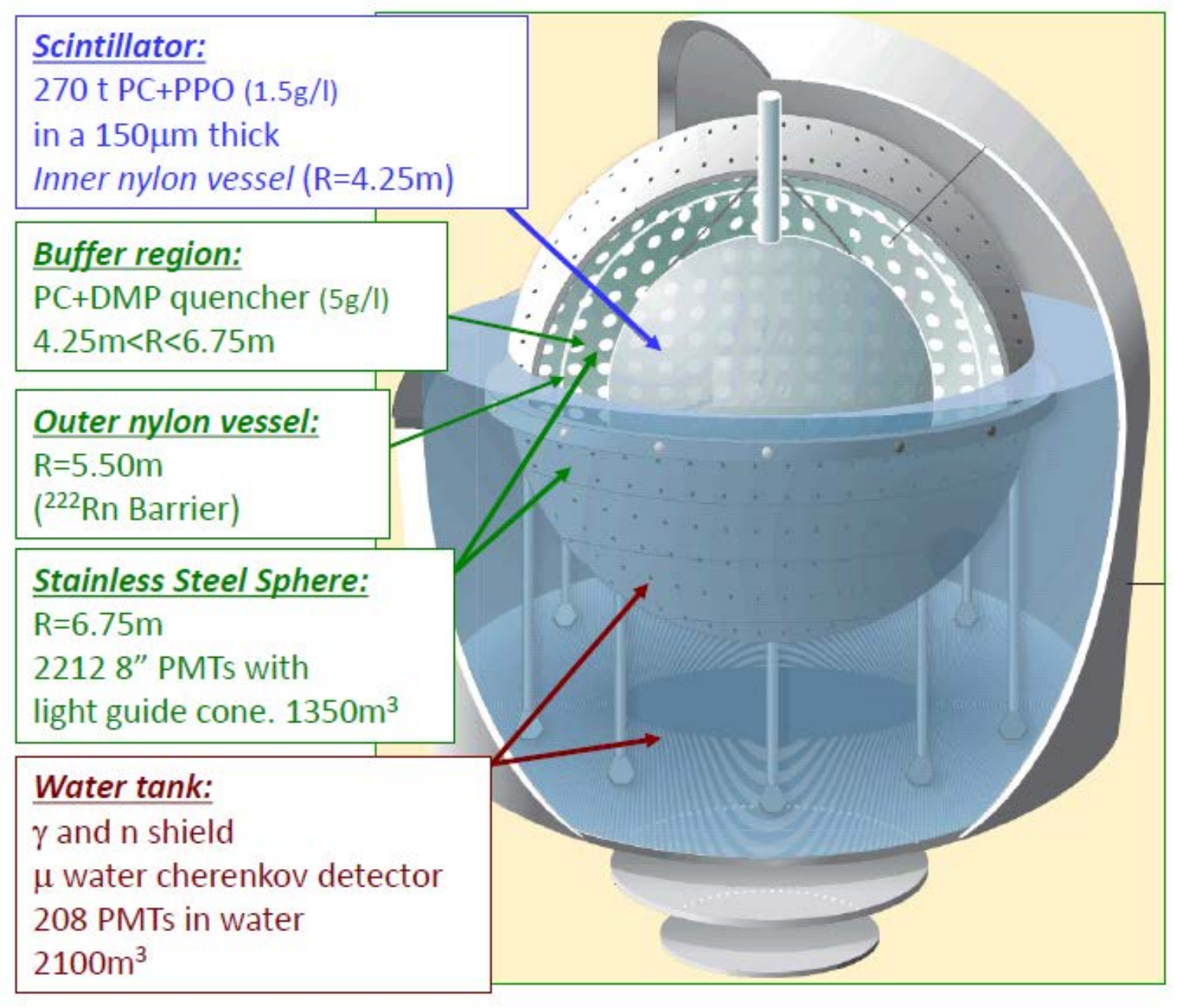}}
  \end{center}
  \caption{Borexino detector}
  \label{fig:det}
 \end{minipage}
\end{figure}

The next nuclear fusion reaction in main sequence stars like the Sun is the
following:
\begin{equation}
  4p \to \alpha + 2e^+ + 2\nu_e + 26.73\mbox{MeV}.
  \label{eq:nuc}
\end{equation}
Neutrinos, generated in the Sun core reach the surface of the Sun almost
immediately ($\sim$2sec) unlike other particles. Therefore, the solar neutrino
measurements directly bring information about the current status of the center
of the Sun. The neutrino generation is realized through the pp-chains and
CNO cycle. The model of the Sun including these reactions is called
the 'Standard Solar Model' (SSM)~\cite{bi:ssm},
and predicts the solar neutrino flux and spectra good accuracy.
Fig.~\ref{fig:nuspc} shows the predicted spectra and
observable energy region for several experiments.
The advantage of Borexino is the capability to measure Solar neutrinos
in elastic scattering, from $\sim$ 0.2 to $\sim$20 MeV.
The wide energy range allowed to measure in real-time both $^8$B and $^7$Be
neutrinos, and pep, CNO and pp neutrinos are also future targets.

The physics motivation of research in solar neutrinos is twofold.
In the neutrino oscillation field, even though the discovery of MSW-LMA
scenario for the last decades,~\cite{bi:sk}~\cite{bi:sno}~\cite{bi:kaml}
the survival probability in $\nu_e$ was very poor constraint before Borexino.
In solar physics, Borexino can help in solving the metallicity controversy
between the new solar composition calculation and helioseismology.
The present goals of Borexino are the measurements of
the precise $^7$Be flux, its day-night asymmetry,
and finally CNO and pp neutrino observation in future.
Other purposes of Borexino are geo-neutrinos~\cite{bi:geo}
and SuperNova neutrinos.

\section{Borexino detector}
\label{sec2}
 The Borexino is an ultra-high radiopure large volume liquid scintillator
detector, located underground (3500m water equivalent) in Gran Sasso in Italy. 
The detector is shown in Fig.~\ref{fig:det}.
The inner core scintillator is a target for neutrino detection, and
consists of 270 tons of pseudocumene as a solvent doped with 1.5 g/l of PPO
as a solute. It is contained in a 4.25m radius of spherical nylon vessel.
The scintillation light are detected by 2212 8-inch photomultiplier tubes
(PMTs) mounted on a stainless steel sphere (SSS).
In order to reduce external $\gamma$ and neutron backgrounds from PMTs and
the rock, the inner scintillator is shielded by 1000 tons of pseudocumene
doped with 5.0 g/l of dimethylphthalate (DMP) as a quencher in buffer region,
and 2000 tons of pure water outside of SSS.
The external water tank is also used to detect the residual cosmic muons
crossing the detector by Cherenkov light.

 Solar neutrinos are detected via elastic scattering on electrons
in liquid scintillator. The advantages of this measurement are high light yield
($\sim$500 photo electrons/MeV), which realizes
low energy threshold ($\sim$250keV) and good energy energy resolution
($\sim 5\%/\sqrt{\mathstrut E/(1MeV)}$),
and a pulse shape discrimination between $\alpha$ and $\beta$ is also possible.
However, there is no way to distinguish neutrino signal and $\beta$ like events
due to radioactivity, therefore, an extreme radiopurity is required.
Thanks to the liquid scintillator purification system the contamination of
$^{238}$U and $^{232}$Th has been removed, reaching a purification level
better than the designed value of 10$^{-16}$ g/g, 
enough to measure not only $^7$Be solar neutrino, but also 8B and potentially
pep and CNO. More details are reported in~\cite{bi:purify}.

 The trigger condition for an event is 25 hit PMTs within 99 ns time window.
When the detector is triggered, hit time and charge information in a 16$\mu$s
gate are recorded. Event position is reconstructed by comparing the hit time
after the time-of-flight subtraction, with a reference {\it pdf} curve.
Energy is determined by number of hit PMTs or summed their photoelectrons.
These qualities were confirmed by several detector calibrations
discussed in the next section.

\section{Detector calibration}
\label{sec3}
Several internal sources of calibration were inserted in the detector in 2009,
aimed to reduce systematic uncertainties, and to tune the reconstruction
algorithm and Monte Carlo simulation.
In the previous result~\cite{bi:bx1}, energy and position calibrations relied
on internal contaminants such as $^{14}$C, $^{222}$Rn.
The correspondent systematic error on the $^7$Be solar neutrino flux was
at the level of 6\% for both the fiducial volume and the energy scale.
The calibration strategy is based on several sources, alphas, betas, gammas,
and neutrons, at different energies, and in hundreds of insertion positions.
In order to avoid additional background contaminations into the detector,
the source vials were carefully
developed. We use a one inch diameter of quartz sphere for filling radioactive
source such as $^{222}$Rn loaded scintillator or $\gamma$ emitters in aqueous
solution. This quartz sphere was attached to a set of stainless steel bars
with a movable arm which could locate the source in various positions inside
the detector. The nominal position of the source was determined independently
by a system of 7 CCD cameras, whose precision was less than 2cm.

\begin{figure}[hbp]
\begin{center}
\resizebox{0.6\textwidth}{!}{\includegraphics{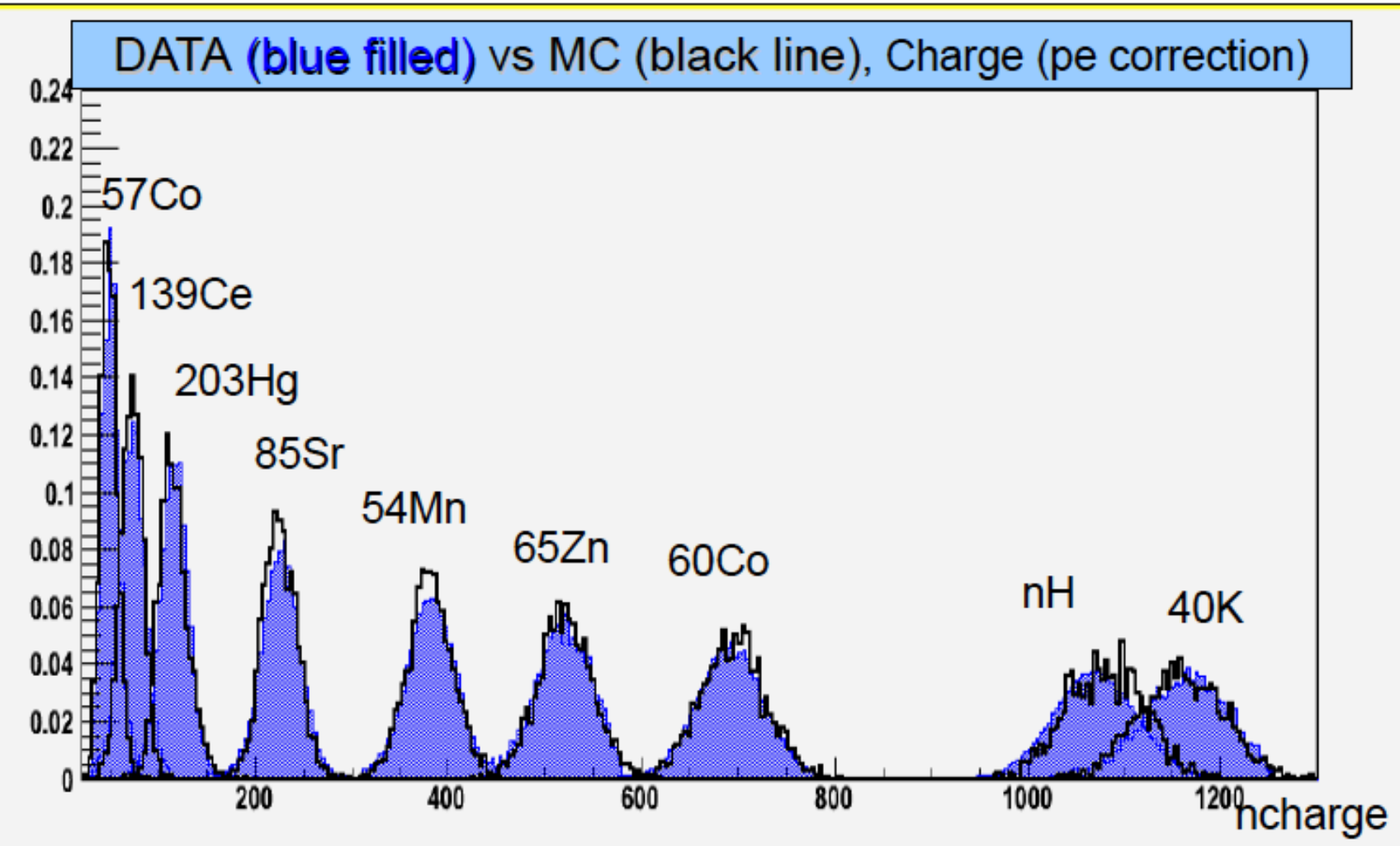}}
\end{center}
\caption{Total photo electron distribution of various $\gamma$ sources between calibration data and Monte Carlo.}
\label{fig:ene}
\end{figure}

 For studying the position reconstruction,
$\alpha$ and $\beta$ events from $^{222}$Rn were used. Comparing the
reconstructed position with nominal position in 184 points of data,
the inaccuracy on the position is less than 3cm level, which is equivalent to
a systematic error of 1.3\% for the overall fiducial volume
in $^7$Be solar neutrino energy region.
The energy response was studied with 8 $\gamma$ sources and Am-Be
neutron source, (2.2MeV $\gamma$ is generated when thermal neutron is captured
by proton.) Fig.~\ref{fig:ene} shows the comparison between calibration data
and Monte Carlo at several energies within the energy region in solar neutrino
analysis. Thanks to this study, the energy scale uncertainty was determined
to be less than 1.5\%. The PMT hit timing in Monte Carlo was tuned by $\alpha$
and $\beta$ events from $^{222}$Rn calibration data. After this tuning,
the particle identification has good agreement between data and Monte Carlo
both $\alpha$ and $\beta$.

\section{Results and implications}
\label{sec4}
\subsection{Data analysis}
\begin{figure}[htp]
\begin{center}
\includegraphics[scale=0.9]{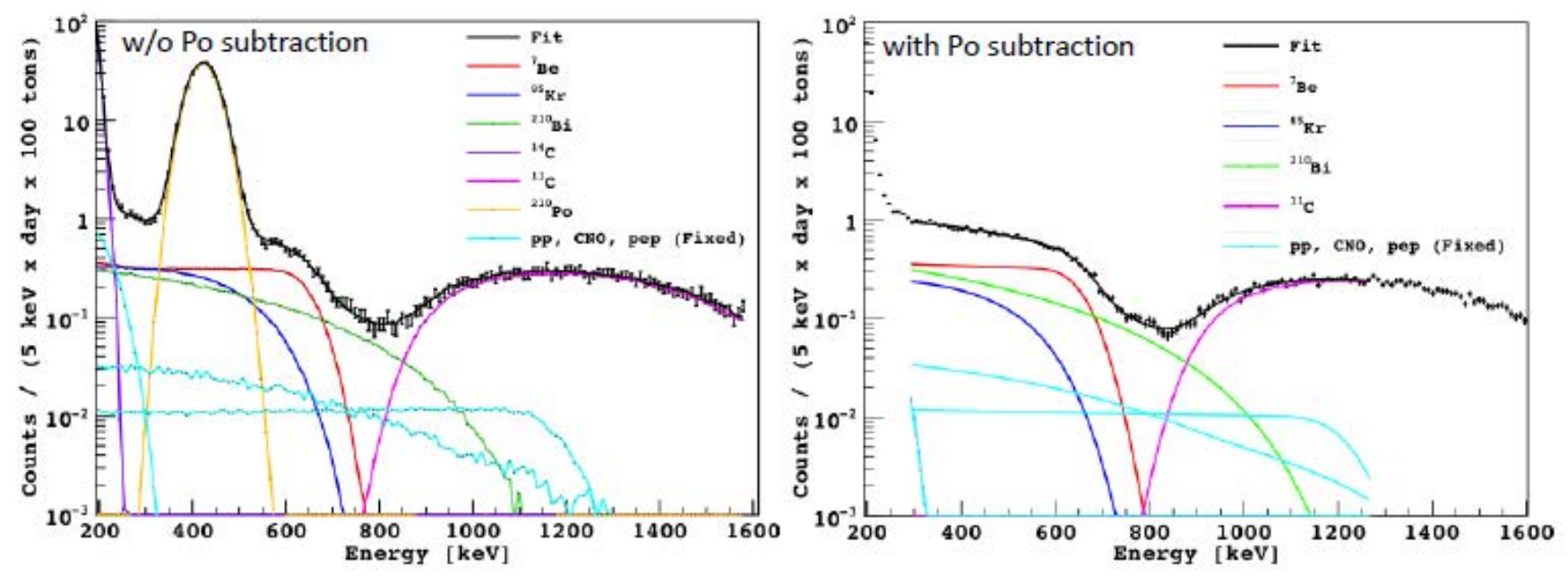}
\end{center}
\caption{Spectrum after analysis cuts, before(left) and after(right) statistically subtraction of $\alpha$s from $^{210}$Po.}
\label{fig:spec}
\end{figure}

The analyzed data set is 740.66 days taken in the period from May 16, 2007
to May 8, 2010. The following selections have been applied;
\begin{itemize}
 \item[1.] Muons and their daughters are rejected. The selection of muons is the combination of the inner and outer detector information. The daughters are defined as all events within 300 ms after each muon.
 \item[2.] A fiducial volume cut was applied to reject the external background events. The reconstructed position must be within a spherical volume of 3m, and also the event position in vertical coordinate must be within $\pm$1.7m to remove background near the poles of the nylon vessel.
\end{itemize}
Finally, the fiducial exposure in this analysis is equivalent to
153.62 ton$\cdot$year.
The left in Fig.~\ref{fig:spec} shows the spectrum after the above reduction.
The remaining peak around 450 keV comes from $\alpha$ events from $^{210}$Po.
For extraction of the $^7$Be solar neutrino signal,
the spectral fit was applied assuming all the intrinsic background components
such as $^{85}$Kr, $^{210}$Bi, $^{14}$C, $^{11}$C.
As for the peak related to $^{210}$Po events,
both fits with and without alpha subtraction were performed.

\subsection{$^7$Be solar neutrino rate}
 The $^7$Be solar neutrino rate was evaluated with the spectral fit
in $46.0\pm1.5(stat.)\pm1.3(sys.)$ counts/day/100ton. 
Total uncertainty including systematic uncertainty is 4.3\%,
(the component of systematic error is 2.7\%)
which is lower than in the previous result:
$\sim$12\%~\cite{bi:bx1}.

\begin{figure}[ht]
 \def\@captype{table}
 \begin{minipage}[t]{0.4\textwidth}
  \begin{center}
   \begin{tabular}{|c|l|c|}
   \hline
   Oscillation & Metal & Rate\\
   \hline
   \hline
   No & High & 74$\pm$4 \\
   \hline
   No & Low  & 67$\pm$4 \\
   \hline
   MSW-LMA & High & 48$\pm$4 \\
   \hline
   MSW-LMA & Low  & 44$\pm$4 \\
   \hline
   \end{tabular}
  \end{center}
  \tblcaption{Expected event rate in Borexino (count/day/100ton) from several hypothesis.}
  \label{tab:expect}
 \end{minipage}
 \hfill
 \begin{minipage}[c]{0.6\textwidth}
  \begin{center}
  \resizebox{0.9\textwidth}{!}{\includegraphics{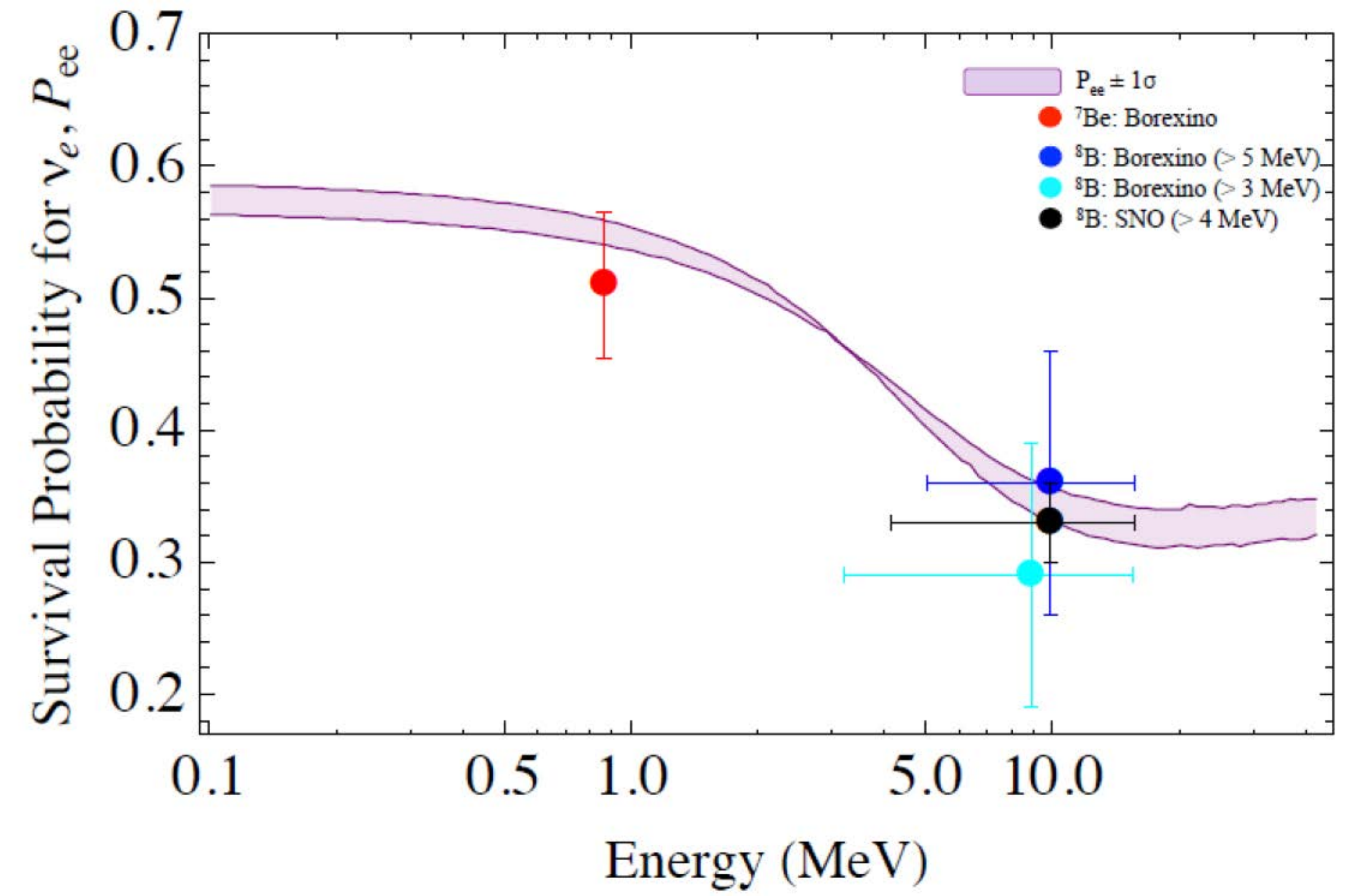}}
  \end{center}
  \caption{Electron neutrino survival probability of expected under the assumption of MSW-LMA scenario and experimental results.}
  \label{fig:surv}
 \end{minipage}
\end{figure}

 Table~\ref{tab:expect} shows the expected rate with several assumptions
both for neutrino oscillation and solar metallicity. Comparing the result
to the expected, no oscillation can be rejected in any metallicity hypothesis.
Fig.~\ref{fig:surv} shows the electron neutrino survival probability
for the $^7$Be and $^8$B~\cite{bi:bxb8} solar neutrino from the Borexino data.
This is the first measurement probing both in the vacuum and in the matter
enhanced regimes combined to $^8$B solar neutrino flux measurement
from the single detector, and the result is good agreement with MSW-LMA
scenario.

\subsection{Day/Night asymmetry}
\begin{figure}[htp]
\begin{center}
\includegraphics[scale=0.9]{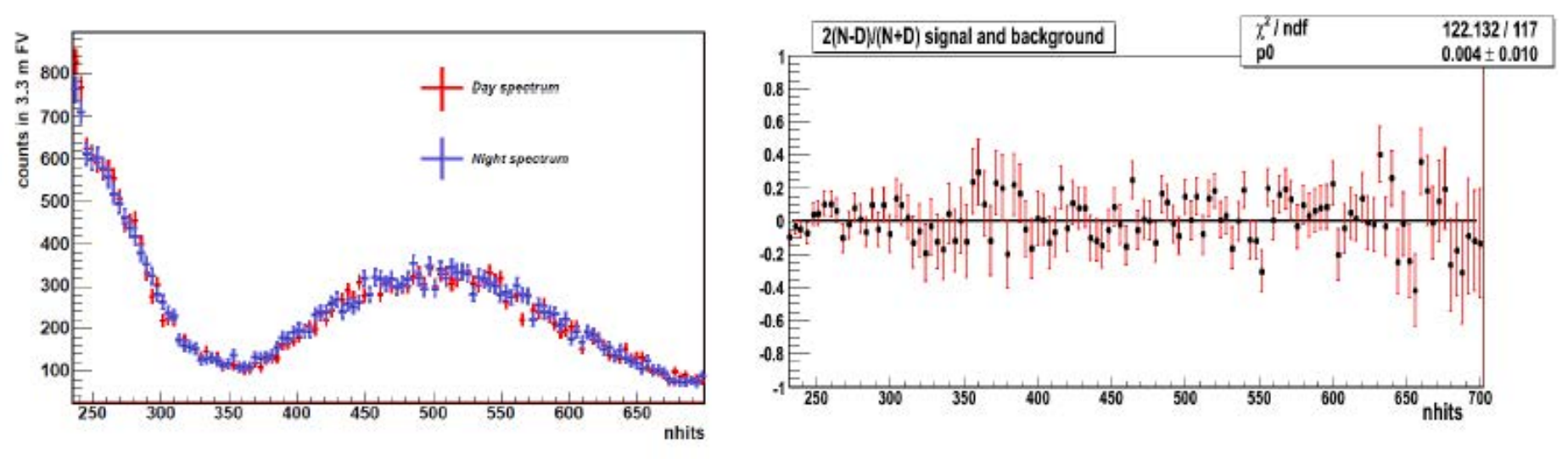}
\end{center}
\caption{(left) Spectrum in day-time and night-time. (right) Day-night asymmetry as a function of energy.}
\label{fig:daynight}
\end{figure}

 In the $^7$Be solar neutrino energy region, the day-night flux difference
is sensitive to distinguish between MSW-LMA and MSW-LOW model,
because about 20\% difference should appear in MSW-LOW region
while no effect in MSW-LMA region.
Fig.~\ref{fig:daynight} show the spectrum both for day-time(D) and
night-time(N), and the energy dependence of its asymmetry which is defined by
(N-D)/((N+D)/2). No significant day-night effect was found, and the overall
asymmetry is 0.7\%. Detailed analysis is now in progress.

\section{Conclusion and perspective}
\label{sec5}
 The Borexino is running since 2007.
The  calibration with radioactive source was performed in 2009.
Increased statistics and calibration lead to a drastic reduction of the
overall error.
The results of $^7$Be solar neutrino, its rate and day-night asymmetry,
and probing both in the vacuum and in the matter enhanced regimes
combined to $^8$B solar neutrino flux measurement, strongly support
an MSW-LMA scenario.

 In order to reduce the internal background and observe pep and CNO
solar neutrinos in near future, the purification with water extraction
is in progress. This measurement will be crucial to distinguish between
high and low metallicity in the solar model.
The measurement of pp solar neutrinos, which is more than 99\% ratio,
is also one of important goal for Borexino, since it promises a complete
understanding the solar interior.

\end{document}